\documentclass[aps,prd,preprintnumbers,showpacs]{revtex4}
\usepackage[dvips]{graphicx}
\begin{document}

%
%

\preprint{Nisho-12/3}
\title{ A Model of Vanishing Cosmological Constant  }
\author{Aiichi Iwazaki} 
\affiliation{International Politics and Economics, Nishogakusha University, \\
6-16 3-bantyo Chiyoda Tokyo 102-8336, Japan.} 
\date{Jun. 25, 2012}

\begin{abstract}
We propose a model in which there exists a real scalar field $q$ satisfying
a condition $\dot{q} =MH$ and its energy density
is given by $\frac{1}{2}\dot{q}^2+V(q)$,
where $H$ is the Hubble parameter ( $H=\dot{a}/a$ ) 
and $M$ is a mass scale characterizing the field. 
We show that the potential $V(q)$ of the field is uniquely determined by the condition. 
The potential depends on the energy densities of background matters.
We find
that the vacuum energy of the matters is cancelled by
the potential of the field. As a result,
the minimum of  
the total energy density of the matters and the field vanishes and is located
at the infinite scale factor $a=\infty$. This remarkable property results without a supersymmetry. 
We show that
the present tiny dark energy 
is caused by early inflation, while the energy is
comparable to Planck scale before the inflation. 
As our model is reduced to
the $\Lambda$CDM model in the limit $M\to 0$,
it is a natural generalization of the $\Lambda$CDM model.
\end{abstract}	
\hspace*{0.3cm}
\hspace*{1cm}

\maketitle

\section{introduction}

Recent observations\cite{1} have strongly suggested 
the acceleration of the cosmic expansion at present.
The acceleration would be caused by
the cosmological constant or a fluid with the equation of state $\omega\simeq -1$.
It is an intriguing idea that the cosmological constant is
not constant but varying with time.
Hence, much attentions\cite{2} have recently been paid to such fluids describing 
the dynamical cosmological constant. There is, however, no guiding principle
for the construction of a promising model of the dynamical cosmological constant. 
For instance, in models using scalar fields describing the cosmological constant,  
various potentials of the fields have been proposed, but they are plagued with
many free parameters.

Models of dynamical cosmological constant must satisfy observational constraints\cite{cons}, but
the constraints are not sufficiently stringent so as to uniquely determine a promising model.
Furthermore, there are serious problems to be solved. One of the problems is why
the vacuum energy density $\sim (10^{-3}\mbox{eV})^4 $ at present is extremely smaller than 
the natural scale $\sim (\mbox{Planck mass})^4$. 
The other one is why the cosmic expansion begins to accelerate nearly at the present epoch.
At least, the expansion of the Universe does not accelerate 
around the epoch of
the recombination. It is a coincidence that we
observe the acceleration of the cosmic expansion at the present epoch.
The solutions to these problems could not be found so far in previous models,
although the coincident problem is alleviated in models with tracking fields\cite{3}. 
( So called "tracking condition" for the potentials of the scalar fields can be a guiding principle 
for the construction of 
models. But there are a variety of potentials\cite{potential,dexp} satisfying the condition. Thus, it is
not restrictive so that a promising model is uniquely determined. Furthermore, the vacuum energy
is forced to vanish by tuning parameters in these models. 
Thus, the cosmological constant problem is not solved. )

In this paper
we propose a guiding principle for the construction of a specific model of dynamical cosmological constant.
The principle is that there exists a real scalar field $q$ satisfying a condition $\dot{q}=MH$
where $H$ is the Hubble parameter ( $=\dot{a}/a$ ) and $M$ is a mass scale characterizing the field.
We also assume that its energy density is given by $\frac{1}{2}\dot{q}^2+V(q)$. 
Then, we can uniquely determine the potential $V(q)$ of the field. It is consisted of two parts,
an intrinsic part and a part depending on the energy densities of background matters. 
The intrinsic potential is an ordinary potential depending only on the field,
while the other potential depends on the energy densities of the matters.
This latter property gives rise to 
a remarkable property that the minimum of the total energy density
( the energy densities of $q$ and the matters ) vanishes.
Namely, 
the vaccum energy vanishes as if there is a supersymmetry.

In the next section \ref{2} we explain our assumptions as well as our guiding principle
for the construction of dynamical cosmological constant.
We derive the potential of the field $q$ by using the principle.
In the section \ref{3} we show
that the total energy density of matters and the field vanishes at $a=\infty$.
Namely, the vacuum energy vanishes.
In section \ref{4} we analyze the dynamical properties of the field $q$. 
In section \ref{5} we discuss observational constraints on parameters in our model.
As a result we find that only one parameter $M$ remains as a free parameter.
We also find that the model is reduced to the $\Lambda$CDM in the limit of $M\to 0$.
In section \ref{6} we discuss that the small dark energy at present is caused by
inflation.
In section \ref{7} we discuss how the cosmological constant and coincidence problems
are alleviated in our model.
In section \ref{8} we discuss our principle and summarize our results.

\section{our model}
\label{2}

We consider a spatial homogeneous flat space Universe with the scale factor $a(t)$;
$ds^2=dt^2-a(t)^2d\vec{x}^2$. We suppose that a real scalar field $q$ exists as well as
matters ( radiations or non-relativistic matter ) in the Universe.
Now we explain our assumptions.
The first of all, 
the field obeys the field equation,

\begin{equation}
\label{eq}
\ddot{q}+3H\dot{q}+\partial_q V=0
\end{equation}
with the Hubble parameter $H=\dot{a}/a$,
where $V$ denotes the potential of the field. 
( The dot denotes a derivative in $t$ ).
The potential is determined by
a condition we propose in this paper.  
The condition is given in the following,

\begin{equation}
\label{p}
\frac{\dot{q}}{M}=\frac{\dot{a}}{a}=H,
\end{equation}
where $M$ denotes a mass scale characterizing the field.
Namely, the scalar field characterizes the expansion rate of the Universe. 
Furthermore, we assume an Einstein equation with the gravitational constant $G$ such that

\begin{equation}
\label{ein}
H^2=\frac{8\pi G(\rho_m+\rho_q)}{3}=\frac{\rho_m+\rho_q}{m^2}
\end{equation}
with $m^2\equiv 3/8\pi G$,
as well as the continuity equation

\begin{equation}
\label{con}
\dot{\rho}_m=-3H(\rho_m+p_m),
\end{equation}
where $\rho_m$ ( $p_m$ ) denotes the energy density ( pressure ) of the matters.
Here we assume that the energy density $\rho_q$ of the field $q$ is given by
$\rho_q\equiv \dot{q}^2/2+V$. If we impose the continuity equation on the field $q$,
that is, $\dot{\rho}_q=-3H(\rho_q+p_q)$ with $p_q\equiv \dot{q}^2/2-V$,
it gives rise to the field equation (\ref{eq}). Thus it is not independent of the equation (\ref{eq}).

Without the condition eq(\ref{p}), the field is a component in the Universe 
and do not play any special roles. But as we show below, 
with the condition the potential $V$ of the field
must depend on the energy density of the matters. 
In the sense
the field is a very special component in the Universe.


\vspace*{0.3cm}
Now, we derive the potential $V$ of the scalar field $q$ by using the condition in eq(\ref{p}).
First, we rewrite the Einstein equation in(\ref{ein}) such that

\begin{equation}  
H^2=\frac{\rho_q+\rho_m}{m^2}=\frac{\dot{q}^2/2+V+\rho_m }{m^2}=\frac{M^2H^2/2+V+\rho_m}{m^2}.
\end{equation}

Thus, we can express the Hubble parameter such that  

\begin{equation}
\label{H}
H^2=\frac{2(V+\rho_m)}{2m^2-M^2},
\end{equation}
where it is natural to take $2m^2-M^2>0$; the mass $M$ characterizing the field $q$
is less than the Planck scale $\sim m$.
 
Furthermore, 
using the field equation $\ddot{q}+3H\dot{q}+\partial_qV=M\dot{H}+3MH^2+\partial_qV=0$,
we obtain

\begin{equation}
\dot{q}(M\dot{H}+3MH^2+\partial_qV)=\frac{d}{dt}(\frac{M^2H^2}{2}+V)+3M^2H^3=0,
\end{equation}
where we used the condition $\dot{q}=MH$.

Inserting the Hubble parameter in eq(\ref{H}) into this equation, 
we obtain

\begin{equation}
\label{VV}
2a\frac{d}{da}V+6\lambda V=-\lambda( a\frac{d}{da}\rho_m+6\rho_m ),
\end{equation}	
with $\lambda\equiv M^2/m^2$,
where we used the relation $d/dt=\dot{a}\,d/da$. 

This is the equation to determine the potential $V$ when 
the energy density $\rho_m$ of the matters is given as a function of $a$. 
For example in the non-relativistic matter ( we simply call it as matter ) dominated period 
when $\rho_m$ is given such as $\rho_m\equiv \rho_0a^{-3}$,
we find 

\begin{equation}
\label{vm}
V=\frac{\lambda \rho_m}{2(1-\lambda)}+V_0a^{-3\lambda}
\quad \mbox{in the matter dominated period}
\end{equation}
where $V_0$ is a constant determined by the present values of the 
density parameters $\Omega_q$ and $\Omega_m$
; $\Omega_q=\rho_q/(\rho_q+\rho_m)$ and $\Omega_m=\rho_m/(\rho_q+\rho_m)$.
Similarly, we obtain the potential in the radiation dominated period;
$V=\lambda \rho_m/(4-3\lambda)+V_0a^{-3\lambda}$ where $\rho_m\propto a^{-4}$.
In general the solution is given by

\begin{equation}
\label{GV}
V=\frac{\lambda}{2}\Big(-\rho_m+
a^{-3\lambda}\int da \,(3\lambda-6)\,a^{3\lambda-1}\rho_m(a)\Big)+V_0a^{-3\lambda}.
\end{equation}

In this way, using our condition in eq(\ref{p}), we can determine the potential of the field $q$.
We should stress that the potential is consisted of two terms; one is the term 
depending on the energy density of the background matters and the other is the intrinsic
term $V_{\rm{int}}\equiv V_0a^{-3\lambda}$.

\section{zero vacuum energy}
\label{3}

In general, we have zero point vacuum energies in quantum field theories.
They are infinite so that we need proper renormalization.
For example in the standard model of quarks and leptons,
we appropriately renormalize the vacuum energy.
Because its absolute value is not observable 
in the Euclidean space without gravity, 
we may take its value arbitrary. ( The relative value of the vacuum energy is observable, 
e.g. Casimir effect. )
But, the vacuum energy is observable when we take account of gravity. We cannot take
it arbitrary. Thus, we may renormalize the vacuum energy such that
it is just given by the cosmological constant observed at present.
The vacuum energy density ( $\sim (10^{-3}\rm{eV})^4$ )
determined in this way is fairly small compared with the other scales in
the standard model with gravity. 
If there is a supersymmetry, 
the vacuum energy exactly vanishes. but the symmetry is broken at
the scale $\Lambda$ beyond $1$ TeV. 
Then, the energy density would be of the order of $\Lambda^4$,
which is much larger than the observed one $\sim (10^{-3}\rm{eV})^4$. 
We do not know why such large discrepancy is present.
The difficulty of the question comes partly from the fact that 
we have no mechanism controlling the vacuum energy.

\vspace{0.2cm}
Our condition postulated in this paper can give a mechanism for rendering
the vacuum energy to vanish. Indeed,
we can see that the solution $V(a)$ vanishes as $a\to \infty$ 
if $\rho_m \to 0$ as $a\to \infty$.
Therefore,the Hubble parameter $H$ or $V+\rho_m$ vanishes in the limit $a\to \infty$.
It implies that the total energy density of the matter and the field
vanishes in the limit. Because $H^2 \ge 0$ and $\dot{H}<0$ ( as we show later ), 
the minimum of the total energy density is located at $a=\infty$.
Even if we add a constant $c$ to $\rho_m$,
the solution $V(a)$ in eq(\ref{vm}) or eq(\ref{GV}) has a term $-c$ so that the total energy density
proportional to $V+\rho_m$ does not change. Thus, the minimum
remains to vanish. This is a general result caused by the condition in eq(\ref{p}). 
Therefore, the condition leads to the vanishing vacuum energy.

\vspace{0.2cm}
For clarity
we give an example of a classical real scalar field as a background matter.
We consider a massless scalar field whose energy density is given by
$\rho_m\equiv\dot{\phi}^2/2+v$, where $v$ is a constant. The equation of motion
of the field,
$\ddot{\phi}+3H\dot{\phi}=0$,
can be easily solved. The solution is given such that $\dot{\phi}=ca^{-3}$
where $c$ is a constant.
Thus, the energy density $\rho_m$ is given by $c^2a^{-6}/2+v$.
Because the energy density is obtained as a function of $a$,
the potential $V$ of the field $q$ can be found from eq(\ref{GV}) such that
$V=-v+V_0a^{-3\lambda}$.  Therefore, we have the total energy density 
$\simeq V+\rho_m=c^2a^{-6}/2+V_0a^{-3\lambda}$, which
has no constant term proportional to $v$ and vanishes at $a=\infty$.
We find that the total vacuum energy of the matter field $\phi$ and the field $q$
vanishes, even if the vacuum energy of the field $\phi$ takes any value,

\vspace{0.2cm}
Previous models 
of dynamical cosmological constant have been plagued by
the ambiguity of the vacuum energy.
That is, as we cannot properly treat the zero point energies in field theories
without supersymmetry,
there is the ambiguity of the precise value of the vacuum energies in $\rho_m$.
But there is no such ambiguity in our model. As long as our principle holds,
the total energy density should vanish in the vacuum even if we do not know the correct 
vacuum energy of the matter fields.
This is a prominent property in our model.

\section{$\rm{q}$ field}
\label{4}

The potential $V$ in eq(\ref{vm}) was obtained as a function of the scale factor $a$, not the field $q$.
But it is easy to rewrite $V$ in terms of $q$.
Because we explicitly obtain $q=M\log (a/a_0)$ from the condition eq(\ref{p}), 
the potential $V$ can be rewritten such that

\begin{equation}
\bar{V}(q)\equiv V(a=a_0\exp(q/M))=\frac{\lambda \rho_0a_0^{-3}\exp(-\frac{3q}{M})}{2(1-\lambda)}
+V_0a_0^{-3\lambda}\exp(-\lambda \frac{3q}{M})
\quad \mbox{in the matter dominated period},
\end{equation}
where $a_0$ is an integration constant and represents the period when
the field $q$ vanishes.

The potential $\propto\exp(-3\lambda q/M)$ intrinsic for $q$
should be of the order of $M^4$. This is because when $q$ takes the value of $M$,
the potential should be the order of $M^4$.
( As we see later, our model 
is reduced to the $\Lambda$CDM model in the limit of $\lambda \to 0$. 
The parameter $\lambda$ is found much smaller than unity by observations. ) 
Thus, we may assume $V_0a_0^{-3\lambda}=M^4$. 
Then, it follows that

\begin{equation}
\label{V}
\bar{V}(q)=\frac{\lambda \bar{\rho}_0\exp(-\frac{3q}{M})}{2(1-\lambda)}
+M^4\exp(-\lambda \frac{3q}{M}) \quad \mbox{in matter dominated period},
\end{equation}
with $\bar{\rho}_0\equiv \rho_0a_0^{-3}$.


\vspace*{0.3cm}
In this way we can derive the explicit form of the potential $\bar{V}(q)$ from the condition in eq(\ref{p}). 
The potential is composed of two terms; 
one is the term proportional to $\rho_m=\bar{\rho}_0\exp(-\frac{3q}{M})$ ( the first term in eq(\ref{V}) ) and 
the other one $V_{\rm{int}}\equiv M^4\exp(-3\lambda q/M)$ 
is the term intrinsic for the field $q$ ( the second term in eq(\ref{V}) ). 
The former depending on $\rho_m$ is much smaller than $\rho_m$ itself 
because $\lambda \ll 1$
( see later ).
We should stress that the intrinsic term $V_{\rm{int}}(a)\propto a^{-3\lambda}$ 
of the potential decreases more slowly than
the first term $\propto \rho_m(a)\propto a^{-3}$ 
as $a$ increases.  
Therefore, the total energy density $\rho_q(a)+\rho_m(a)$ eventually approaches $V_{\rm{int}}(a)$
with the expansion of the Universe. That is, the field $q$ dominates the Universe.
Thus, we may regard the present value $V_{\rm{int}}(a=1)=V_0$ as the dark energy observed.

We note that the equation of the state of the field $q$ is given by

\begin{equation}
\omega_q=\frac{p_q}{\rho_q}=\frac{\frac{\dot{q}^2}{2}-V}{\frac{\dot{q}^2}{2}+V}
=\frac{\lambda \rho_m+2(\lambda-1)V}{\lambda \rho_m+(\lambda+2)V}\simeq 
\frac{\lambda \rho_m-2V}{\lambda \rho_m+2V}
\end{equation}
with $\lambda \ll 1$,
where we used eq(\ref{p}) and eq(\ref{H}). Therefore, we find that
when the field $q$ dominates the Universe, the expansion accelerates
because $\lambda\rho_m$ is smaller than $V$ so that $\omega_q\simeq -1$.
We mention that with small $\lambda \ll 1$,
the potential in eq(\ref{V}) satisfies "tracking condition"\cite{trac} 
$\bar{V}''(q)\bar{V}(q)/\bar{V}'(q)^2>1$, where
$\bar{V}'\equiv \partial_q\bar{V}$ and $\bar{V}''\equiv \partial_q^2\bar{V}$.

The potential in eq(\ref{V}) has been discussed in previous papers \cite{dexp}, in which
more general ones than the potential $\bar{V}(q)$ have been considered 
such that $A\exp(-\alpha q/M)+B\exp(-\beta q/M)$. The parameters $A,B,\alpha$, $\beta$ and $M$ have been
partially determined so as to obtain a quintessence model with favorable "tracking fields".
But free parameters still remain in these potentials\cite{mexp},
although the parameters are partially fixed by observations. 
Other models\cite{potential} with various potentials have been proposed, but
they are also plagued with
many free parameters.
On the other hand,
there is only one free parameter in our model. Actually,
we have three parameters $M$, $\rho_0$ and $V_0$.
Among them,
$\rho_0$ and $V_0$ are determined by the present values of 
the density parameters $\Omega_q$ and $\Omega_m$. The remaining one $M$
is only one adjustable parameter, which can be 
determined by observing the equation of state of the field $q$ today.
Although our model involves only one adjustable parameter $M$ or $\lambda$, 
the model is shown to be consistent with current observations
as long as we have small $\lambda \,\,(\,\,\ll 1$ ).

We should mention a peculiar property of the field $q$.
In general, when we solve a field equation like eq(\ref{eq}),
we need to impose initial conditions. They may be arbitrarily taken.
But this is not true for the field $q$.
An initial condition for $\dot{q}_{\rm{in}}$ at $t=t_{\rm{in}}$ is determined by
the condition $\dot{q}_{\rm{in}}=MH_{\rm{in}}$ in eq(\ref{p}) where $H_{\rm{in}}$ is 
Hubble parameter at $t=t_{\rm{in}}$. On the other hand, an initial condition
for $q_{\rm{in}}$ at $t=t_{\rm{in}}$ can be arbitrarily taken. 
The arbitrary choice of a value $q_{\rm{in}}$
corresponds to determining the value of $a_0$; $q_{\rm{in}}=M\log(a_{\rm{in}}/a_0)$.
In this way, the dynamics of the field $q$ is fairly restrictive
compared with ordinary scalar fields in the other models of quintessence.
This restriction comes from the condition in eq(\ref{p}).
It claims that the field $q$ is not independent of the scale factor $a$,
although the field equation of $q$ is required to hold.

We can see the actual behavior of the field $q$ by examining the equation of motion,

\begin{equation}
\label{eqq}
\ddot{q}=-\partial_q\bar{V}(q)-3H\dot{q}=
\frac{3\lambda \bar{\rho}_0\exp(-\frac{3q}{M})}{2M(1-\lambda)}
+\frac{3\lambda M^4\exp(-\frac{3\lambda q}{M})}{M}     
-\frac{3\dot{q}^2}{M},
\end{equation}
where the last term in the right hand side acts 
as a friction.
Therefore, a particle with the coordinate $q$ slowly goes down the slope of the potential,
being affected by the friction; the coordinate $q$ logarithmically increases 
with time because $q=M\exp(a/a_0)$ and $a\propto t^{2/3}$ in the matter dominated period.
The initial condition $\dot{q}_{\rm{in}}=MH_{\rm{in}}>0$
implies that the particle never goes up the slope;
it must always goes down. 
This implies that the Universe always expands as long as
it expands at the initial period $t=t_{\rm{in}}$. 
Although the result holds in the matter dominated period,
the similar result is obtained in the radiation dominated period.

It is important to see that 
$\ddot{q}<0$, because the right hand side of the equation can be rewritten
in the following,

\begin{equation}
\ddot{q}=\frac{3}{M}\Big(\frac{\lambda \rho_m}{2(1-\lambda)}+\lambda V_{\rm{int}}
-\frac{2\lambda(V+\rho_m)}{2-\lambda}\Big)
=\frac{3}{M}\Big(-\frac{\lambda \rho_m}{2(1-\lambda)}
-\frac{\lambda^2V_{\rm{int}}}{2-\lambda}\Big)<0,
\end{equation}
with $\lambda \ll 1$,
where we used $\dot{q}^2/2=\lambda (V+\rho_m)/(2-\lambda)$ and 
$V=\lambda\rho_m/2(1-\lambda)+V_{\rm{int}}$ in the matter dominated period.
The same inequality holds in the radiation dominated period.
Hence, the Hubble parameter $H=\dot{q}/M$ always decreases with time; $\dot{H}<0$.
In other words the total energy density always decreases with time.
Thus, its minimum ( $=0$ ) is located at $a=\infty$. That is, 
the vacuum energy vanishes in our model.

It is also instructive to see that the coordinate $q$
linearly increases with time in the period of inflation.
This is because the scale factor $a$ exponentially increases in the period.  
In this way, the solution $q$ of the equation 
represents the expansion history of the Universe.

\section{observational constraints}
\label{5}

We wish to determine the parameters in the potential $V$ by 
examining observational constraints.
First we consider 
the density parameters $\Omega_q$, $\Omega_m$ 
and the equation of state $\omega_q$ in the matter dominated period. They are
given by respectively,

\begin{eqnarray}
\label{omega-q}
\Omega_q&=&\frac{2V+\lambda\rho_m}{2(V+\rho_m)}=\frac{2(1-\lambda)V_0 a^{-3\lambda}
+(2-\lambda)\lambda\rho_m}{2(1-\lambda)V_0 a^{-3\lambda}+(2-\lambda)\rho_m}
\simeq \frac{V_0 a^{-3\lambda}+\lambda\rho_0a^{-3}}{V_0 a^{-3\lambda}+\rho_0a^{-3}},\\
\Omega_m&=&\frac{(2-\lambda)\rho_m}{2(V+\rho_m)}=\frac{(2-\lambda)(1-\lambda)\rho_m}
{2(1-\lambda)V_0a^{-3\lambda}+(2-\lambda)\rho_m}
\simeq \frac{\rho_0 a^{-3}}{V_0 a^{-3\lambda}+\rho_0 a^{-3}},\\
\omega_q&=&\frac{p_q}{\rho_q}=\frac{\frac{\dot{q}^2}{2}-V}{\frac{\dot{q}^2}{2}+V}
=\frac{-2(1-\lambda)^2V_0 a^{-3\lambda}}{2(1-\lambda)V_0 a^{-3\lambda}+\lambda(2-\lambda)\rho_m}
\simeq \frac{-V_0 a^{-3\lambda}}{V_0 a^{-3\lambda}+\lambda\rho_0a^{-3}},
\end{eqnarray}
where we used $\dot{q}^2/2=\lambda (V+\rho_m)/(2-\lambda)$ and took the limit $\lambda \ll 1$.

Using the present values $\Omega^0_q=0.75$ and $\Omega^0_m=0.25$ at $a=1$ ( here we take
$a=1$ as the scale factor today ), we obtain the parameters
$V_0=0.75\rho^0_c$ and $\rho_0=0.25\rho^0_c$ where 
$\rho^0_c$ denotes the critical density $\rho^0_c\equiv \rho_0+V_0$ at present.
Furthermore, we find that 

\begin{equation}
\label{omega}
\omega_q(a\simeq 1)\simeq -1+\frac{4\lambda}{3}+\lambda (1-a), 
\end{equation} 
where we have taken the limit $a\to 1$.
Current observations suggests 
$\omega_q(a=1)\lesssim -0.9$. It leads to the constraint $\lambda\lesssim 0.08$.
As our formula has only one parameter $\lambda$, 
we can check the validity\cite{cons} of our model by observing
$\omega_q(a\simeq 1)$ in detail.

We can examine whether or not the field $q$ affects Big Bang Nucleosynthesis,
which occurs at about $a=10^{-9}$. \\
( As radiations dominates the Universe at the period,
$\rho_m$ in the above formulas eq(\ref{omega-q}) is formally 
replaced by $\rho_m/2$ where $\rho_m$ is the energy density
of radiations proportional to $a^{-4}$. )
Because $\Omega_q\simeq \lambda$ for sufficiently small $a\ll 1$ 
such as $\lambda\, a^{-4}\gg a^{-3\lambda}$,
the field does not affect the Nucleosynthesis\cite{cons,n}
if $\lambda<0.1$. Therefore, our model satisfies
both of the observational constraints with the small parameter $\lambda <0.1$.

\vspace{0.2cm}
We should mention that when we take the limit $M\to 0$ ( $\lambda\to 0$ ), 
the field $\dot{q}=MH$ vanishes and
the total energy density of the Universe becomes $\rho_0a^{-3}+V_0$ in the matter
dominated period; $\dot{q}^2/2+V(a)\to V_0$ in the limit $M\to 0$. 
Thus, our model
is reduced to the standard $\Lambda$CDM model with the cosmological constant $\Lambda=3V_0/m^2$.
( Here we take the limit 
keeping the critical density today $\rho^0_c=\rho_0+V_0$ fixed. )
Because the standard $\Lambda$CDM model is consistent with current observations,
our model is also consistent with them as long as the parameter $\lambda=M^2/m^2 $ is much small.

It is interesting to see that the energy density of the field 
dominates the Universe only around the present
epoch $a=1$. In order to see it we calculate

\begin{equation}
\label{ra}
\frac{\Omega_q}{\Omega_m}=\frac{2(1-\lambda)V_0 a^{-3\lambda}
+(2-\lambda)\lambda\rho_m}{(2-\lambda)(1-\lambda)\rho_m}\simeq 
\frac{V_0 a^{-3\lambda}
+\lambda\rho_m}{\rho_m}=\frac{\frac{V_0}{\rho_0} a^{-3\lambda}
+\lambda a^{-3}}{a^{-3}}=\lambda+\frac{V_0}{\rho_0}a^{3(1-\lambda)}\simeq \lambda+3a^3
\end{equation}
with $\lambda\ll 1$, 
where we used the present values $\Omega^0_q=3\Omega^0_m$ 
or $V_0=3\rho_0$. Thus, we find that the field $q$ dominates the Universe 
just after the period $a=(1/3)^{1/3}\simeq 0.7$.

More precisely we estimate the turning point from the deceleration to the acceleration of the Universe, that is, 
the period when $\ddot{a}=0$.
The period is obtained by solving the equation, 

\begin{eqnarray}
\frac{\ddot{a}}{a}&=&-\frac{\rho_m+\rho_q+3p_q}{2m^2}=
\frac{2(V+\rho_m)+3\lambda (V+\rho_m)-3V(2-\lambda )}{2m^2(2-\lambda )} \nonumber \\
&=&\frac{(-4+6\lambda)V+(2+3\lambda)\rho_m}{2m^2(2-\lambda )}\simeq 
\frac{(\rho_0(2+3\lambda)+V_0(-4+6\lambda)a^{3-3\lambda}-2\rho_0\lambda )a^{-3}}{2m^2(2-\lambda )}=0.
\end{eqnarray}
Thus, it follows that

\begin{equation}
\frac{\ddot{a}}{a}\propto \frac{\rho_0}{V_0}(2+\lambda)+(-4+6\lambda)a^{3-3\lambda}
=\frac{\rho_0}{V_0}(2+\lambda)+(-4+6\lambda)\frac{(1+z)^{3\lambda}}{(1+z)^3}=0,
\end{equation}
with $a=1/(1+z)$.
As $\rho_0/V_0=1/3$ derived from the present density parameters,
we find that the turning point from
the deceleration to the acceleration is given by

\begin{equation}
\ddot{a}(z)=0 \quad \mbox{at} \quad z\simeq 0.8 \,\,\,(\, 0.82 \,)
\quad \mbox{for} \quad \lambda=0.1 \,\,\,(\, \lambda=0.01 \,),
\end{equation}
while $\ddot{a}=0$ at $z(a)\simeq 0.82$ in the $\Lambda$CDM model obtained 
by taking the limit $\lambda=M^2/m^2\to 0$.

\vspace*{0.3cm}
We have shown that using the condition
 in eq(\ref{p}) we uniquely determine 
the model of the dynamical cosmological constant. Our model involves only one free parameter $\lambda$
and is consistent with the current observations when $\lambda <0.1$. 
We have also shown that the model is reduced to
the $\Lambda$CDM in the limit $\lambda \to 0$.
Therefore, our model is a natural extension of the standard $\Lambda$CDM.  

\section{inflation}
\label{6}

We now proceed to show why the present value of dark energy 
$V_{\rm{in}}(a=1)=V_0=M^4a_0^{3\lambda}$ is much smaller than
the natural scale $m^4$ of the Universe. 
In other words, we show
why $a_0$ is extremely small such as
$a_0=(V_0/M^4)^{1/3\lambda}\sim 10^{-400}$ 
with $V_0=(10^{-3}\rm{eV})^4$ and $\lambda\sim 0.1$.
The period at $a=a_0$ is the epoch when the field $q$ vanishes; $q=M\log(a/a_0)$.
Without inflation the scale factor $a$ at most
smoothly increases such as $t^n$ with positive $n$ being $O(1)$. It never reaches the value $a=1$
within the life time of the Universe.
Thus, we suppose that the period $a=a_0$ 
is earlier than the beginning of inflation\cite{inf}. Then, by the inflation
the scale factor $a$ exponentially increases
and can reaches $a=1$.
The inflation 
makes the dark energy  $V_{\rm{in}}\propto a^{-3\lambda}$
exponentially small,  
while the potential $V_{\rm{in}}$ takes the large value $M^4
\simeq \lambda^2 m^4$ at $a=a_0$
before the inflation. Thus, 
the present small dark energy is realized, that is, 
$V_{\rm{in}}(a)=V_0a^{-3\lambda}=M^4(a_0/a)^{3\lambda}\to V_0$ as $a\to 1$.

In order to show how it can be actually realized in our model, 
we may explicitly take $\rho_{\rm{inf}}=ka^{\epsilon}$ as a flat potential of inflaton\cite{inf}
with $0<|\epsilon| \ll 1$. 
Then, $\rho_{\rm{inf}}$ is almost constant in $a$ ( $ka_i^{\epsilon}$ is a constant
of the order of $M'^4$, in which
the inflation is supposed to begins at $a=a_i$ and $M'$ is a typical scale of inflaton; $M'<M$. )
The precise form of the inflaton potential is not necessary for the later discussion.
Only what we need is the flatness of the potential in $a$. 
So, the parameter $\epsilon$ must be extremely small. We also assume that $k\epsilon$ is positive
in order for the total energy density to be positive, see later. 
( We need extremely small but non vanishing parameter $\epsilon$, because
if $V_{\rm{inf}}\propto a^{\epsilon}$ is exactly constant in $a$, i.e. $\epsilon=0$,
the constant $V_{\rm{inf}}$ is cancelled 
by the term $-V_{\rm{inf}}$ in $V$
so that it does not contribute to total energy density. )  

We assume that the kinetic energy of the inflaton is much small compared with the potential energy
when it rolls along the slope of the flat potential. Similarly, the kinetic energy of the field $q$
is small compared with its potential as we have discussed in previous sections.
Therefore, we may 
assume that the total energy density is given by  
the potentials of the field $q$ and the inflaton,
The potential $V(a)$ is obtained by
solving eq(\ref{VV}) with $\rho_m$ replaced by $\rho_{\rm{inf}}$,

\begin{equation}
\label{iV}
V=-\frac{k(1+\frac{\epsilon}{6})}{1+\frac{\epsilon}{3\lambda}}a^{\epsilon}+V_{\rm{in}}
\simeq -ka^{\epsilon}+\frac{k\epsilon a^{\epsilon}}{3\lambda}+V_0a^{-3\lambda}
\end{equation}
where we used that $\epsilon \ll \lambda $. Thus, the Hubble parameter is given by

\begin{equation}
\label{IH}
H^2=\frac{2(V+\rho_{\rm{inf}})}{2m^2-M^2}
\simeq (1+\frac{\lambda}{2})m^{-2}
\Bigl(\frac{k\,\epsilon \,a^{\epsilon}}{3\lambda}+V_0a^{-3\lambda}\Bigr).
\end{equation}
where the first term in the right hand side of the equation 
represents the contribution of the inflaton and the second term
does the contribution of the field $q$.
As we pointed out, the factor $k\epsilon $ must be positive.
As the second term decreases
with the expansion of the Universe much faster than the first term, 
the first term eventually dominates the Hubble parameter $H$. 
The inflation begins at 
$a=a_i \gg a_0$ 
where $a_i$ satisfies 

\begin{equation}
\frac{k\,\epsilon\, a_i^{\epsilon}}{3\lambda}=\frac{M'^4 \epsilon}{3\lambda}  \simeq V_0a_i^{-3\lambda} 
=M^4\Big(\frac{a_0}{a_i}\Big)^{3\lambda}, 
\end{equation}
That is, $a_i\simeq a_0\Big(\frac{M^4\lambda}{M'^4\epsilon}\Big)^{1/3\lambda}$.
Once the inflation starts, the potential $V_{\rm{in}}(a)=M^4(a_0/a)^{3\lambda}$
rapidly decreases with time.
Therefore, we have extremely small potential 
$V_{\rm{in}}(a=a_f)=M^4(a_0/a_f)^{3\lambda}$ at 
the end $a=a_f (\gg a_i$) of the inflation compared with the potential
$V_{\rm{in}}(a=a_i)=M^4(a_0/a_i)^{3\lambda}$ at the beginning $a=a_i$ of the inflation.
This is because the ratio $V_{\rm{in}}(s_i)/V_{\rm{in}}(s_f)=(a_f/a_i)^{3\lambda}$ is exponentially large.
In other words the dark energy can be much small at the end of the inflation,
although it is not necessarily small at the beginning of the inflation.
In this way 
we can obtain much small dark energy at present.


We may estimate e-folding number of the inflation leading to the small dark energy today.
As the present value $V_0$ is of the order of $(10^{-3}\rm{eV})^4$,
the e-folding number $N$ of the inflation may be roughly estimated such that

\begin{equation}
\label{ef}
N=\log\frac{a_f}{a_i}\sim 920, \quad \mbox{because} \quad 
\frac{V_{\rm{in}}(a_i)}{V_{\rm{in}}(a_f=1)}
=\Bigl(\frac{a_i}{a_f=1}\Bigr)^{-3\lambda}=\Bigl(\frac{a_i}{a_f=1}\Bigr)^{-0.3}
\sim \Bigl(\frac{M}{10^{-3}\rm{eV}}\Bigr)^4\sim 10^{120},
\end{equation} 
with $\lambda=0.1$,
where we assumed that the inflation begins at $a_i=a_0$ and ends at $a_f=1$. 
Obviously, this e-folding number is much larger
than the actual one because the inflation begins after the period $a_0$ ( $<a_i$ ) and ends 
much earlier than today $a_f\ll 1$.
The above estimation is an example to show that the inflation really makes the potential much small
by taking an appropriate value of the e-folding number N. 
In this way we can have much small intrinsic potential energy at the end of the inflation.
( The reason why $a_0$ is much small compared with the present value $a=1$
is obvious. Such a small $a$ is exponentially inflated. Then, it smoothly increases to 
reach the present value $a=1$ 
with the standard expansion of the Universe. ) 

\vspace{0.2cm}

We would like to point out the behavior of $q$ in the period of the inflation.
As we obtained the potential $V$ in eq(\ref{iV}), we can easily solve the equation (\ref{eqq}) for $q$.
Inserting $V\simeq -ka^{\epsilon}=-ka_0^{\epsilon}\exp(\epsilon q/M)$ into the equation,
$\ddot{q}+3H\dot{q}+\partial_qV(q)\simeq 3\dot{q}^2/M-k\epsilon a_0^{\epsilon}/M=0$,
we obtain a solution such that $\dot{q}=\sqrt{k\epsilon a_0^{\epsilon}/3}$.
Thus, we find that the field $q$ linearly increases with time.
This implies that the scale factor $a$ exponentially increases with time.


\vspace{0.2cm}

Finally, we point out that it is difficult to realize the above inflation scenario
by means of real scalar fields.
To see it,
we consider a massive real scalar field described by the Lagrangian
$L=\frac{1}{2}(\dot{\phi}^2-\mu^2\phi^2)$.
The equation of motion of the field $\ddot{\phi}+3H\dot{\phi}+\mu^2\phi=0$
can be solved by assuming the slow roll motion of the field, i.e. $\ddot{\phi}\ll \mu^2\phi$
in the period of the inflation when $H$ is a constant given by $h$.
The solution is given by $\phi\propto\exp(-\mu^2t/3h)=(\exp ht)^{(-\mu^2/3h^2)}$.
Thus, we have $\phi=c a^{-\mu^2/3h^2}$ with a constant $c$ because $a\propto \exp(ht)$
in the period of the inflation. 
The energy density of the field
is given such that $\rho_m=ka^{-2\mu^2/3h^2}$ with $k\equiv c^2\mu^2/2$ in term of
the scale factor $a$. 
The slow roll condition $\ddot{\phi}\ll \mu^2\phi$ is satisfied 
when the inequality $\mu^2 \ll h^2$ holds. Namely, we obtain the flat inflaton potential $ka^{\epsilon}$
with $\epsilon\equiv -2\mu^2/3h^2$; $|\epsilon|\ll 1$. It seems apparently that 
the flat potential leads to the exponential increase of the scale factor $a\sim \exp(ht)$,
that is the inflation. But this is not true in our case where 
the square of the Hubble parameter is proportional to $V+\rho_m$, not to $\rho_m$.
We find from eq(\ref{IH}) that
the consistency condition $k\epsilon >0$ must hold. However,
the our model of the inflaton does not satisfy this condition. 
Indeed we have $k\epsilon<0$;
$k=c^2\mu^2/2>0$ and $\epsilon=-2\mu^2/3h^2 <0$. 
In this way, the simple inflation model using the real scalar field does not work.
The difficulty of making naive inflaton model
comes from the absence of the constant term in the total energy density.

\section{cosmological constant and coincidence problems}
\label{7}

We would like to discuss both problems of the cosmological constant and coincidence
in our model.

\vspace{0.2cm}
First we discuss the problem of the cosmological constant.
The problem is why the vacuum energy density ( $V_{\rm{in}}$ in our model )
at present is much small compared with a natural scale $m^4 \sim (\mbox{Planck mass})^4$ 
given by gravity. If we had approximate supersymmetry 
in the standard model, such a small energy density can be explained when the breaking scale of 
the supersymmetry is such small as $10^{-3}$eV.
But the real breaking scale is beyond $1$ TeV.
 
On the other hand, as if there is a supersymmetry in our model, 
the vacuum energy vanishes
owing to the condition eq.(\ref{p}). Namely,
even if there are any quantum corrections of matter fields for vacuum energy,
the corrections are cancelled by the field $q$ so that the total vacuum energy vanishes. 
Owing to this remarkable property, the small vacuum energy density at present
can be realized by the inflation. That is, it tremendously dilutes the
large vacuum energy density of the order of $M^4=m^4\lambda^2$.
Although we still need a tuning of parameters 
to obtain the appropriate small energy density $(10^{-3}\rm{eV})^4$ at present,
the problem of the cosmological constant is fairly alleviated.   
 
\vspace{0.2cm}
The coincidence problem is why
the vacuum energy density is comparable 
to the energy density of the dark matter at present.
In our model this corresponds to the question why $V_0=V_{\rm{in}}(a=1)\sim \rho_m(a=1)=\rho_0$.
In order to discuss the problem we note the equation(\ref{ra}),

\begin{equation}
\frac{\Omega_q}{\Omega_m} \simeq \lambda+ (V_0/\rho_0)a^3 \quad 
\mbox{in the non relativistic matter dominated period}
\end{equation} 
where we took the limit $\lambda \ll 1$.
This implies that for the arbitrarily chosen value $V_0/\rho_0$, 
the equality $\Omega_q\sim \Omega_m$
arises at the epoch $a=(\rho_0/V_0)^{1/3}$. After the epoch,
the field $q$ dominates the Universe.
As explained above, the inflation tremendously dilutes the vacuum energy density $V_{\rm{in}}$.
After the inflation, $V_{\rm{in}}$ ( $\propto a^{-3\lambda}$ ) much slowly decreases 
with the expansion of the Universe because $\lambda \ll 1$. On the other hand,
the energy density of the matter $\rho_m\propto a^{-3}$ or $a^{-4}$ 
decreases much faster than $V_{\rm{in}}$.  
Therefore, the coincidence problem can be described as 
how the inflation dilutes the vacuum energy density such that 
the ratio $V_0/\rho_0$ is of the order of unity.

In this way both solutions for the cosmological constant and coincidence problems 
lie in the detail of inflation model: How does the inflation brings out 
extremely small vacuum energy $V_0\simeq (10^{-3}\rm{eV})^4$ and renders
the ratio $V_0/\rho_0$ of the order of unity ?

\section{summary and discussion}
\label{8}

To summarize, postulating the presence of the real scalar field $q$
satisfying the condition $\dot{q}/M=H$,
we have derived the simple analytical model of the dynamical cosmological constant.
The model has several remarkable properties. 
The most remarkable one is the property of the zero vacuum energy.
Because the potential of the field $q$ depends on the energy densities of background matters,
even if any quantum corrections of the matters are added to the vacuum energy, 
they are cancelled by the potential of the field $q$. In addition, the total energy
density decreases with the expansion of the Universe and vanishes at $a= \infty$.
Therefore, the vacuum energy vanishes even if any ordinary matters are present in the Universe.

Furthermore, we have shown that inflation tremendously decreases the dark energy density $V_{\rm{in}}$
because $V_{\rm{in}}\propto a^{-3\lambda}$. Thus, 
the inflation can make the energy density extremely small,
while
it is comparable to the scale 
$m^4$ or $M^4=\lambda^2 m^2$ before
the inflation.
As a result the dark energy density becomes much small such as $(10^{-3}\rm{eV})^4$ at present.
The coincidence problem is addressed by the question how the inflation
gives rise to the ratio of the order of unity
between the dark energy $V_{\rm{in}}(a=1)$ and the dark matter $\rho_m(a=1)$, that is,
$V_{\rm{in}}(a=1)/\rho_m(a=1)=V_0/\rho_0\sim O(1)$.
Hence, to answer the problem, we need to examine the detail of the inflation mechanism.  
As there is only one adjustable parameter $\lambda$ involved in our model,
we can check the model 
by observing the equation of state $\omega_q(a\sim 1)$ at present.
Our model is reduced to the $\Lambda$CDM model in the limit $\lambda\to 0$.
Thus, our model is a natural generalization of the $\Lambda$CDM model.

We have proposed the relation $\dot{q}=MH$ between the field $q$ and
the Hubble parameter $H$. But we do not know how the relation can be
incorporated in a Lagrangian formalism. Furthermore,
we do not know generalized equations describing the spatially inhomogenuity of $q$.
Probably, there would be a symmetry behind our model,
which guarantees the vanishing vacuum energy. If the Lagrangian formalism 
is obtained, we will be able to find the symmetry as well as the equations describing
the spatially inhomogenuity of the field $q$. 
In our model, 
we assume the relation in addition to
the ordinary equations, i.e. Einstein equations with matters
and the field equation of $q$. Thus,   
It seems apparently that our model is consistent with the Einstein gravity.
But the relation could not be incorporated in the Einstein gravity. 
Therefore,
our model would be a key for pursuing more fundamental
theory of gravity, which has the property of the vanishing vacuum energy
without supersymmetry.



\end{document}